\documentclass{imsart}

\usepackage{algorithm,algorithmic,amsfonts,amssymb,amsmath,amscd,amsthm,bm,latexsym}
\usepackage{natbib}
\usepackage{graphicx}
\usepackage{picins}
\usepackage{upgreek}
\usepackage{units}

\newcommand\GB{\mathfrak{B}}
\newcommand\E{\mathbb{E}}

\newcommand\bx{\mathbf{x}}
\newcommand\by{\mathbf{y}}
\newcommand\bz{\mathbf{z}}
\newcommand\beps{\boldsymbol{\epsilon}}
\newcommand\bX{\mathbf{X}}

\newcommand\dd{\text{d}}
\newcommand{\piabc}{\pi_\text{ABC}}

\newcommand{\bs}{\mathbf{s}}
\newcommand{\sobs}{S(\bx^0)}
\renewcommand{\rho}{\varrho}

\newenvironment{example}[1][Example]{\begin{trivlist}
\item[\hskip \labelsep {\bfseries #1}]}{\hfill$\blacktriangleleft$\end{trivlist}}

\begin{document}

\begin{frontmatter}

\title{Bayesian Computational Tools\protect\thanksref{T1}}
\runtitle{Bayesian Computational Tools}
\thankstext{T1}{Christian P. Robert, CEREMADE, Universit{\' e} Paris-Dauphine, 75775 Paris cedex 16, France
{\sf xian@ceremade.dauphine.fr}. Research partly supported by the Agence Nationale de la Recherche (ANR,
212, rue de Bercy 75012 Paris) through the 2012--2015 grant ANR-11-BS01-0010 ``Calibration'' and by a Institut 
Universitaire de France senior chair. C.P.~Robert is also affiliated as a part-time researcher with CREST, INSEE, Paris.}

\begin{aug}
 \author{\snm{Christian P.~Robert}}
 \affiliation{Universit{\'e} Paris-Dauphine, CEREMADE, and CREST, Paris}
\end{aug}

\begin{abstract}
This chapter surveys advances in the field of Bayesian computation over the past twenty years, from a
purely personnal viewpoint, hence containing some ommissions given the spectrum of the field. Monte Carlo,
MCMC and ABC themes are thus covered here, while the rapidly expanding area of particle methods is only
briefly mentioned and different approximative techniques like variational Bayes and linear Bayes methods do
not appear at all. This chapter also contains some novel computational entries on the double-exponential 
model that may be of interest {\em per se}.
\end{abstract}

\end{frontmatter}

\section{Introduction}

It has long been a bane of the Bayesian approach that the solutions it proposed were
intellectually attractive but inapplicable in practice. While some numerical analysis
solutions were suggested \citep[see, e.g.][]{smith:1984}, they were not in par with the
challenges raised by handling non-standard probability densities, especially in high dimensional
problems. This stumbling block in the development of the Bayesian perspective became clear when
new simulations methods appeared in the early 1990's and the number of publications
involving Bayesian methods rised significantly (no test available!). While those methods were on
principle open to any type of inference, they primarily benefited the Bayesian paradigm as they
were ``ideally" suited to the core object of Bayesian inference, namely a mostly intractable posterior
distribution. 

This chapter will not cover the historical developments of computational
methods (see, e.g., \citealp{robert:casella:2011}) nor the technical
implementation details of simulation techniques (see, e.g.,
\citealp{doucet:defreitas:gordon:2001},
\citealp{robert:casella:2004}, \citealp{robert:casella:2009} and
\citealp{brooks:etal:2011}), but instead focus on examples of application of those
methods to Bayesian computational challenges. Given the limited length of the chapter,
it is to be understood as a sequence of illustrations of the main computational
tools, rather than a comprehensive introduction, which is to be found in the
books mentioned above and below.

\section{Some computational challenges}\label{sec:chal}

The starting point of a Bayesian analysis being the posterior distribution, let us recall that it
is defined by the product 
$$
\pi(\theta|x) \propto \pi(\theta)f(x|\theta)
$$ 
where $\theta$ denotes the parameter and $x$ the data. (The symbol $\propto$
means that the functions on both sides of the symbol are proportional as
functions of $\theta$, the missing constant being a function of $x$, $m(x)$.) The
structures of both $\theta$ and $x$ can vary in complexity and dimension,
although we will not discuss the non-parametric case when $\theta$ is infinite
dimensional, referring the reader to \citet{denison:holmes:mallick:smith:2002} for an 
introduction.  The prior distribution is most often available in closed form, being chosen 
by the experimenter, while the likelihood function $f(x|\theta)$ may be too involved to be 
computed even for a given pair $(x,\theta)$. In special cases where $f(x|\theta)$ allows 
for a demarginalisation representation
$$
f(x|\theta) = \int f(x,z|\theta) \,\text{d}z\,,
$$
where $g(x,z|\theta)$ is a (manageable) probability density, we will call $z$ the missing data. However,
the existence of such a representation does not necessarily implies it is of any use in computations. (We
will encounter both cases in Sections \ref{sec:MCMC} and \ref{sec:ABC}.)

Since the posterior distribution is defined by
$$
\pi(\theta|x) = {\pi(\theta)f(x|\theta)}\bigg/{\int_\Uptheta \pi(\theta)f(x|\theta)\,\text{d}\theta}
$$
a first difficulty occurs because of the normalising constant: the denominator is very rarely available
in closed form. This is an issue only to the extent that the posterior density is defined up to a constant.
In cases where the constant does not matter, inference can be easily conducted without the constant. Cases
when the constant matters include testing and model choice, since the marginal likelihood
$$
m(x) = \int_\Uptheta \pi(\theta)f(x|\theta)\,\text{d}\theta
$$
is central to the Bayesian procedures addressing this inferential problem. Indeed, when comparing two models
against the same dataset $x$, the prefered Bayesian solution (see,
e.g., \citealp{robert:2001}, Chapter 5, or \citealp{jeffreys:1939}) is 
to use {\em the Bayes factor}, defined as the ratio of marginal likelihoods
$$
\GB_{12}(x) = \frac{m_1(x)}{m_2(x)} = \dfrac{\int_{\Uptheta_1} \pi(\theta_1)f(x|\theta_1)\,\text{d}\theta_1}
{\int_{\Uptheta_2} \pi(\theta_2)f(x|\theta_2)\,\text{d}\theta_2}\,,
$$
and compared to $1$ to decide which model is most supported by the data (and how much).
Such a tool---quintessential for running a Bayesian test---means that for almost any
inference problem---barring the very special case of conjugate priors--- there
is a computational issue, not the most promising feature for promoting an
inferential method. This aspect has obviously been addressed by the community,
see for instance \citet{chen:shao:ibrahim:2000} that is entirely dedicated to
the problem of approximating normalising constants or ratios of normalising
constants, but I regret the issue is not spelled out much more clearly as one of the major
computational challenges of Bayesian statistics (see also \citealp{marin:robert:2010}).

\begin{example}[Example 1]
As a benchmark, consider the case \citep{marin:pillai:robert:rousseau:2011} 
when a sample $(x_1,\ldots,x_n)$ can be issued
either from a normal $\mathcal{N}(\mu,1)$ distribution or from a double-exponential
$\mathcal{L}(\mu,1/\sqrt{2})$ distribution with density
$$
f_0(x|\mu) = \frac{1}{\sqrt{2}} \exp\{ -\sqrt{2}|x-\mu|\}\,.
$$
(This case was suggested to us by a referee of \citealp{robert:cornuet:marin:pillai:2011},
however I should note that a similar setting opposing a normal model to (simple) exponential data used as a 
benchmark in \cite{ratmann:2009} for ABC algorithms.)
Then, as it happens, the Bayes factor $B_{01}(x_1,\ldots,x_n)$
is available in closed form, since, under a normal 
$\mu\sim\mathcal{N}(0,\sigma^2)$ prior, the marginal likelihood for the normal
model is given by
\begin{align*}
m_1(x_1,\ldots,x_n) &= \int (2\pi)^{-n/2} \prod_{i=1}^n \exp\{ -(x_i-\mu)^2/2\} 
\exp\{-\mu^2/2\sigma^2\}\,\dd\mu/\sqrt{2\pi}\sigma \\
&= (2\pi)^{-n/2} \exp\{ -\sum_{i=1}^n (x_i-{\bar x_n})^2/2\}\\
&\quad\times\int \exp[ -\{ (n+\sigma^{-2})\mu^2 -2n\mu{\bar x_n} 
+ n ({\bar x_n})^2 \}/2 ] \,\dd\mu/\sqrt{2\pi}\sigma \\
&= (2\pi)^{-n/2} \exp\{ -\sum_{i=1}^n (x_i-\bar x_n)^2/2\}\,\\
&\quad \times\exp\{ -n\sigma^{-2} (\bar x_n)^2/2 (n+\sigma^{-2}) \} /\sigma\sqrt{n+\sigma^{-2}}
\end{align*}
and, for the double-exponential model, by (assuming the sample is sorted)
\begin{align*}
m_0(x_1,\ldots,x_n) &=
\int 2^{-n/2} \prod_{i=1}^n \exp\{ -\sqrt{2}|x_i-\mu|\} \exp\{-\mu^2/2\sigma^2\}\,\dd\mu/\sqrt{2\pi}\sigma \\
&= \frac{2^{-n/2}}{\sqrt{2\pi}\sigma} \sum_{i=0}^n 
\int_{x_i}^{x_{i+1}} \prod_{j=1}^i e^{ \sqrt{2}x_j-\sqrt{2}\mu} \prod_{j=i+1}^n e^{ -\sqrt{2}x_j+\sqrt{2}\mu} 
e^{-\mu^2/2\sigma^2}\,\dd\mu \\
&= \frac{2^{-n/2}}{\sqrt{2\pi}\sigma} \sum_{i=0}^n 
\int_{x_i}^{x_{i+1}} e^{ \sqrt{2}\sum_{j=1}^i x_j- \sqrt{2}\sum_{j=i+1}^n x_j+\sqrt{2}(n-2i)\mu} 
e^{-\mu^2/2\sigma^2}\,\dd\mu \\
&= 2^{-n/2} \sum_{i=0}^n  e^{ \sqrt{2}\sum_{j=1}^i x_j- \sqrt{2}\sum_{j=i+1}^n x_j+2(n-2i)^2\sigma^2/2}\\
&\qquad \times \int_{x_i}^{x_{i+1}} e^{-\left\{ \mu-\sqrt{2}(n-2i)\sigma^2 \right\}^2 
/ 2\sigma^2}\,\dd\mu/\sqrt{2\pi}\sigma \\
&= 2^{-n/2} \sum_{i=0}^n  e^{ \sqrt{2}\sum_{j=1}^i x_j- \sqrt{2}\sum_{j=i+1}^n x_j+(n-2i)^2\sigma^2}\\
&\quad \times \left[ \Phi(\{x_{i+1}-\sqrt{2}(n-2i)\sigma^2\}/\sigma) - \Phi(\{x_{i}
-\sqrt{2}(n-2i)\sigma^2\}/\sigma) \right]
\end{align*}
with obvious conventions when $i=0$ ($x_0=-\infty$) and $i=n$ ($x_{n+1}=+\infty$). To illustrate the
consistency of the Bayes factor in this setting, Figure \ref{fig:BF1a} represents the distributions
of the Bayes factors associated with 100 normal and 100 double-exponential samples of sizes 50 and 200, respectively. 
While the smaller samples see much overlay in the repartition of the Bayes factors, for 200 observations, in both
models, the log-Bayes factor distribution concentrates on the proper side of zero, meaning that it 
discriminates correctly between the two distributions for a large enough sample size.
\end{example}
\piccaption{\label{fig:BF1a}
{\small Repartition of the values of the log-Bayes factors associated with 100 normal
(orange) and 100 double-exponential samples (blue) of size 50 (left) and 200 (right), 
estimated by the default R density estimator.}}
\piccaptioninside
\parpic[l]{\includegraphics[width=3cm]{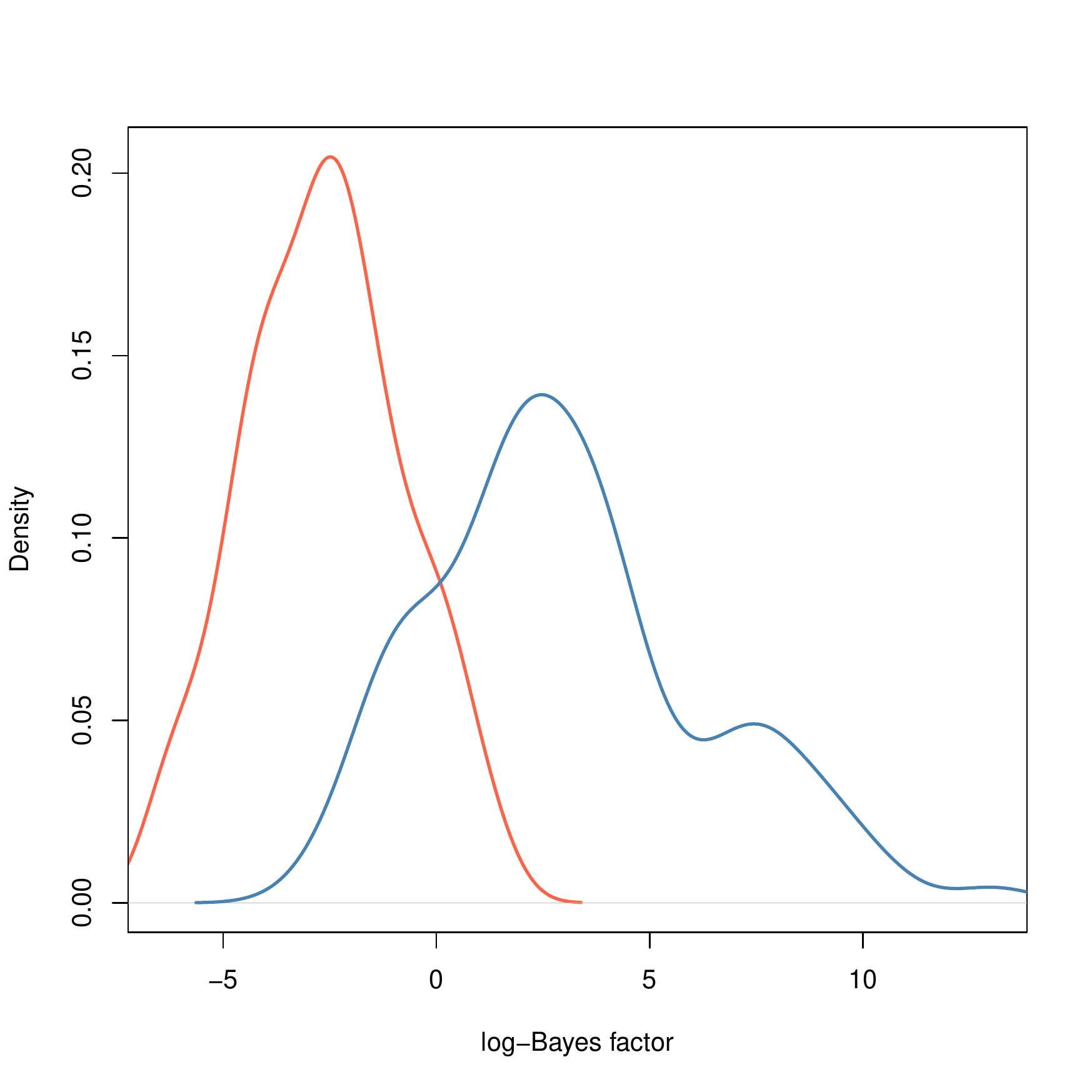} \includegraphics[width=3cm]{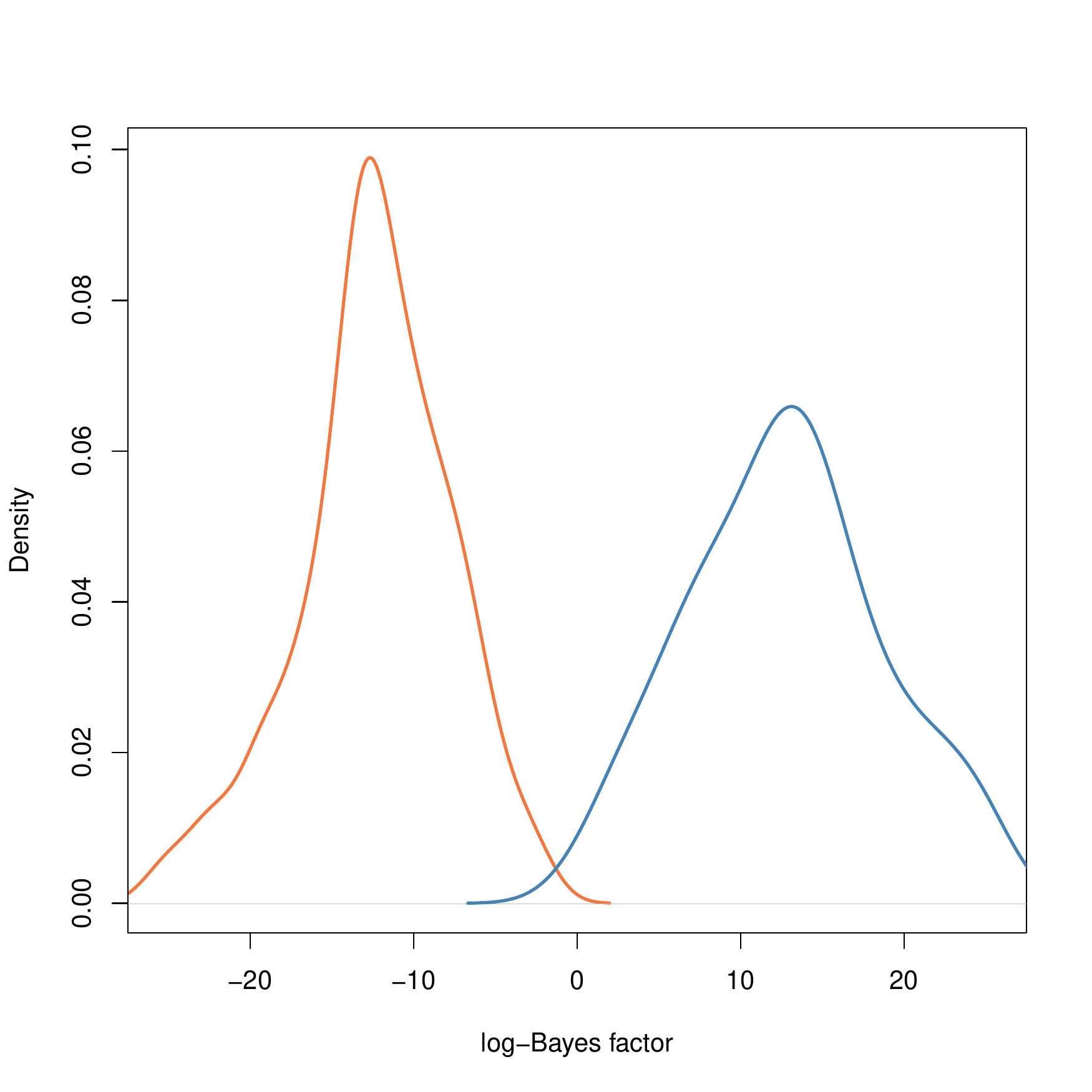}}



Another recurrent difficulty with using posterior distributions for inference is the derivation of
credible sets---the Bayesian version of confidence sets (see, e.g.,
\citealp{robert:2001})---since they are usually defined as highest posterior
density regions:
$$
C_\alpha(x) = \left\{ \theta;\,\pi(\theta|x)\ge \kappa_\alpha(x) \right\}\,,
$$
where the bound $k_\alpha$ is determined by the credibility of the set
$$
\mathbb{P}(\theta\in C_\alpha(x)|x) = \alpha\,.
$$
While the normalisation constant is irrelevant in this problem, determining the collection
of parameter values such that $\pi(\theta)f(x|\theta)\ge \kappa_\alpha(x)$ and calibrating the lower bound
$\kappa_\alpha(x)$ on the product $\pi(\theta)f(x|\theta)$ to achieve proper coverage are non-trivial
problems that require advanced simulation methods. Once again, the issue is somehow overlooked in the
literature.

While one of the major appeals of Bayesian inference is that it is not reduced to an
estimation technique---but on the opposite offers a whole range of inferential
tools to analyse the data against the proposed model---, the computation of
Bayesian estimates is nonetheless certainly one of the better addressed computational
issues. This is especially true for posterior moments like the posterior mean 
$\E^\pi[\theta|x]$ since they are directly represented as ratios of integrals
$$
\E^\pi[\theta|x] = \dfrac{\int_\Uptheta \theta \pi(\theta)f(x|\theta)\,\text{d}\theta}{\int_\Uptheta
\pi(\theta)f(x|\theta)\,\text{d}\theta}\,.
$$
The computational problem may however get involved for several reasons, including for instance
\begin{itemize}
\item[--] the space $\Uptheta$ is not Euclidean and the problem imposes shape constraints (as in
some time series models);
\item[--] the dimension of $\Uptheta$ is large (as in non-parametrics);
\item[--] the estimator is the solution to a fixed point problem (as in the credible
set definition);
\item[--] simulating from $\pi(\theta|x)$ is delicate or even impossible;
\end{itemize}
the latter case being in general the most challenging and thus the most studied,
as the following sections will show.

\section{Monte Carlo methods}\label{sec:MC}

Monte Carlo methods have been introduced by physicists in Los Alamos, namely Ulam, von Neumann, 
Metropolis, and their collaborators in the 1940's (see \citealp{robert:casella:2011}). The idea behind Monte
Carlo is a straightforward application of the {\em law of
large numbers}, namely that, when $x_1,x_2,\ldots$ are i.i.d.~from the distribution $f$, the empirical average
$$
\dfrac{1}{T}\,\sum_{t=1}^T h(x_t)
$$
converges (almost surely) to $\mathbb{E}_f[h(X)]$ when $T$ goes to $+\infty$.
While this perspective sounds too simple to apply to complex problems---either
because the simulation from $f$ itself is intractable or because the variance
of the empirical average is too large to be manageable---, there exist more
advanced exploitations of this result that lead to efficient simulation
solutions. 

\begin{example}[Example 1 (bis)]
Consider computing the Bayes factor 
$$
\mathfrak{B}_{01}(x_1,\ldots,x_n)= m_0(x_1,\ldots,x_n)/m_1(x_1,\ldots,x_n)
$$
by simulating a sample
$(\mu_1,\ldots,\mu_T)$ from the prior distribution, $\mathcal{N}(0,\sigma^2)$.
The approximation to the Bayes factor is then provided by
$$
\widehat{\mathfrak{B}_{01}} =
\sum_{t=1}^T \prod_{i=1}^n f_0(x_i|\mu_t) \bigg/ \sum_{t=1}^T \prod_{i=1}^n f_1(x_i|\mu_t)\,,
$$
given that in this special case the {\em same} prior and the {\em same} Monte Carlo
samples can be used. Figure \ref{fig:BF2} shows the convergence of $\widehat{\mathfrak{B}_{01}}$
over $T=10^5$ iterations, along with the true value. The method exhibits convergence.
\end{example}
\piccaption{\label{fig:BF2}
{\small Convergence of a Monte Carlo approximation of $\mathfrak{B}_{01}(x_1,\ldots,x_n)$
for a normal sample of size $n=19$, along with the true value (dash line).}}
\piccaptioninside
\parpic[l]{\includegraphics[width=5cm]{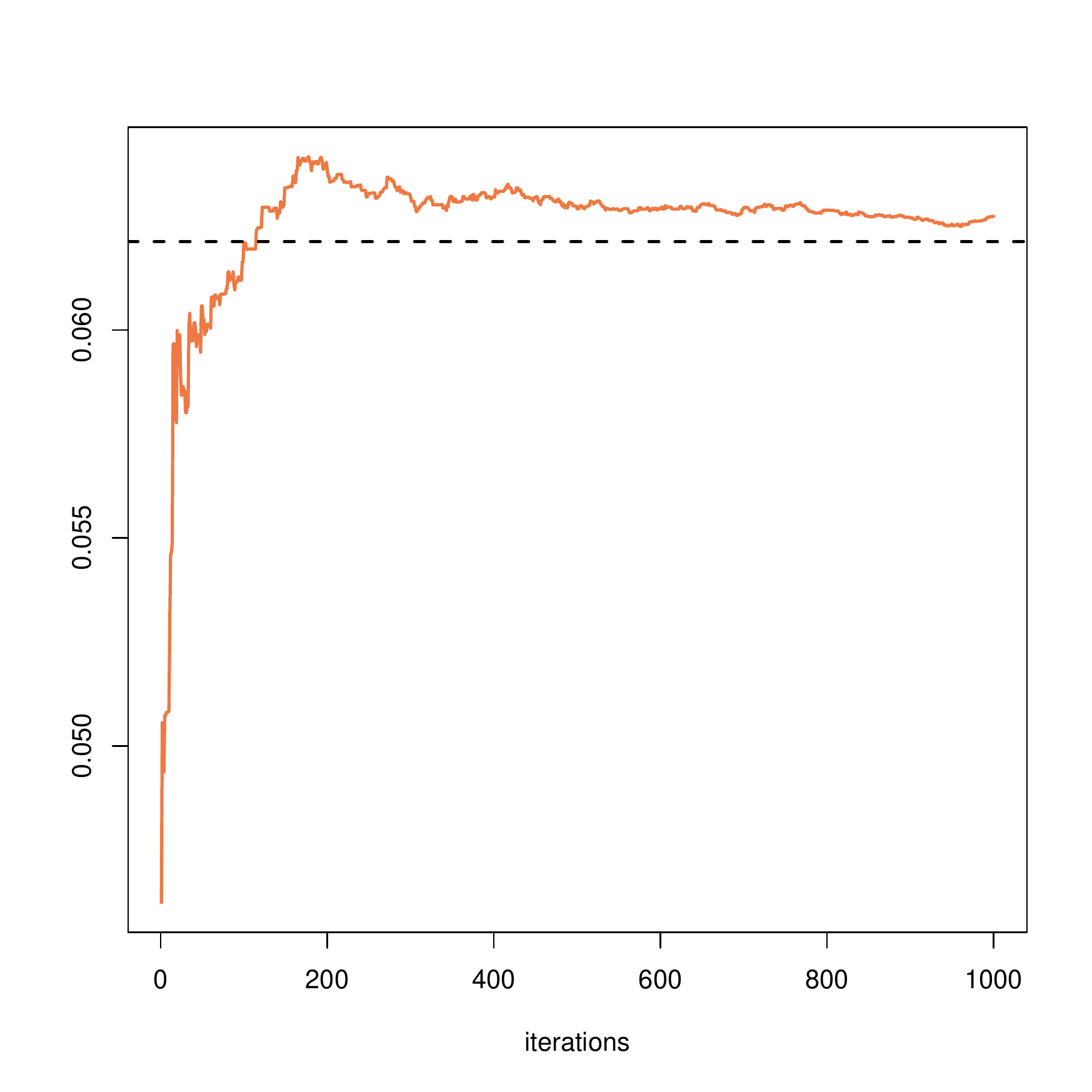}}

The above example can also be interpreted as an illustration of importance sampling, in
the sense that the prior distribution is used as an importance function in both integrals.
We recall that importance sampling is a Monte Carlo method where the quantity of interest
$\E_f[h(X)]$ is expressed in terms of an expectation under the importance density $g$,
$$
\E_f[h(X)] = \E_g[h(X)f(X)/g(X)]\,,
$$
which allows for the use of Monte Carlo samples distributed from $g$.
Although importance sampling is at the source of the particle method \citep{doucet:defreitas:gordon:2001},
I will not develop this useful sequential method any further, but instead
briefly introduce the notion of bridge
sampling \citep{meng:wong:1996} as it applies to the approximation of Bayes factors 
\begin{align*}
\mathfrak{B}_{01} (x) &= 
\int_{\Uptheta_0} f_0(x|\theta_0) \pi_1(\theta_0) \,\text{d}\theta_0\\
&\quad  \bigg/ \int _{\Uptheta_1}f_1(x|\theta_1) \pi_1(\theta_1) \,\text{d}\theta_1
\end{align*}
(and to other ratios of integrals). This method handles
the approximation of ratios of integrals over identical spaces (a severe constraint), 
by reweighting two samples from both posteriors, through a well-behaved type of harmonic average.

More specifically, when $\Uptheta_0=\Uptheta_1$, possibly after a reparameterisation of
both models to endow $\theta$ with the same meaning, we have 
\begin{eqnarray*}
\mathfrak{B}_{01} (x) &=& {\displaystyle\int_{\Uptheta_0} f_0(x|\theta) \pi_0(\theta) \alpha(\theta) {\pi}_1(\theta|x) 
\text{d}\theta }\bigg/ 
{\displaystyle \int_{\Uptheta_1} f_1(x|\theta) \pi_1(\theta) \alpha(\theta) {\pi}_0(\theta|x) \text{d}\theta } \\  
&\approx& 
\dfrac{{n_1}^{-1} \sum_{j=1}^{n_1} f_0(x|\theta_{1,j}) \pi_0(\theta_{1,j}) \alpha(\theta_{1,j})}
{{n_0}^{-1} \sum_{j=1}^{n_0} f_1(x|\theta_{0,j}) \pi_1(\theta_{0,j}) \alpha(\theta_{0,j})}
\end{eqnarray*}
where $\theta_{0,1},\ldots,\theta_{0,n_0}$ and $\theta_{1,1},\ldots,\theta_{1,n_1}$
are two independent samples coming from
the posterior distributions $\pi_0(\theta|x)$ and $\pi_1(\theta|x)$,
respectively. (This identity holds for any function $\alpha$ guaranteeing the
integrability of the products.) However, there exists a quasi-optimal
solution, as provided by \citet{gelman:meng:1998}:
$$
{\alpha^\star(\theta) \propto \dfrac{1}{n_0{\pi}_0(\theta|x) + n_1  {\pi}_1(\theta|x)}} \,.
$$
While this optimum cannot be used---given that it relies on the normalising constants
of both $\pi_0(\cdot|x)$ and $\pi_1(\cdot|x)$---, a practical implication of the
result resorts to an iterative construction of $\alpha^\star$. 
We gave in \citet{chopin:robert:2010} an alternative representation of the bridge factor that
bypasses this difficulty (if difficulty there is!).

\begin{example}[Example 1 (ter)]
If we want to apply the bridge sampling solution to the normal versus double-exponential
example, we need to simulate from the posterior distributions in both models. The normal
posterior distribution on $\mu$ is a normal $\mathcal{N}(n\bar x_n/(n+\sigma^{-2}),1/(n+\sigma^{-2}))$
distribution, while the double-exponential distribution can be derived as a mixture of $(n+1)$ truncated normal
distributions, following the same track as with the computation of the marginal distribution above. The
sum obtained in the above expression of $m_0(x_1,\ldots,x_n)$ suggests interpreting $\pi_0(\mu|x_1,\ldots,x_n)$ 
as (once again assuming $\bx$ sorted)
$$
\sum_{i=0}^n \omega_i \mathcal{N}^\text{T}(\sqrt{2}(n-2i)\sigma^2,\sigma^2,x_i,x_{i+1})
$$
where $\mathcal{N}^\text{T}(\delta,\tau^2,\alpha,\beta)$ denotes a truncated normal distribution,
that is, the normal $\mathcal{N}(\delta,\tau^2)$ distribution restricted to the interval $(\alpha,\beta)$,
and where the weights $\omega_i$ are proportional to those summed in $m_0(x_1,\ldots,x_n)$ (see Example 1 (bis)).
The outcome of one such simulation is shown in Figure \ref{fig:Lapost} along with the target density: as seen there,
since the true posterior can be plotted against the histogram, the fit is quite acceptable.
If we start with an arbitrary estimation of $\mathfrak{B}_{01}$ like $\mathfrak{b}_{01}=1$, successive iterations
produce the following values for the estimation: $11.13$, $10.82$, $10.82$, based on $10^4$ samples from each 
posterior distribution (to compare with an exact ratio equal
to $10.3716$ and a Monte Carlo approximation of $10.55$).
\end{example}
\piccaption{\label{fig:Lapost}
{\small Histogram of $10^4$ simulations from the posterior distribution associated with a double-exponential
sample of size 150, along with the curve of the posterior (dashed lines).}}
\piccaptioninside
\parpic[r]{\includegraphics[width=6cm]{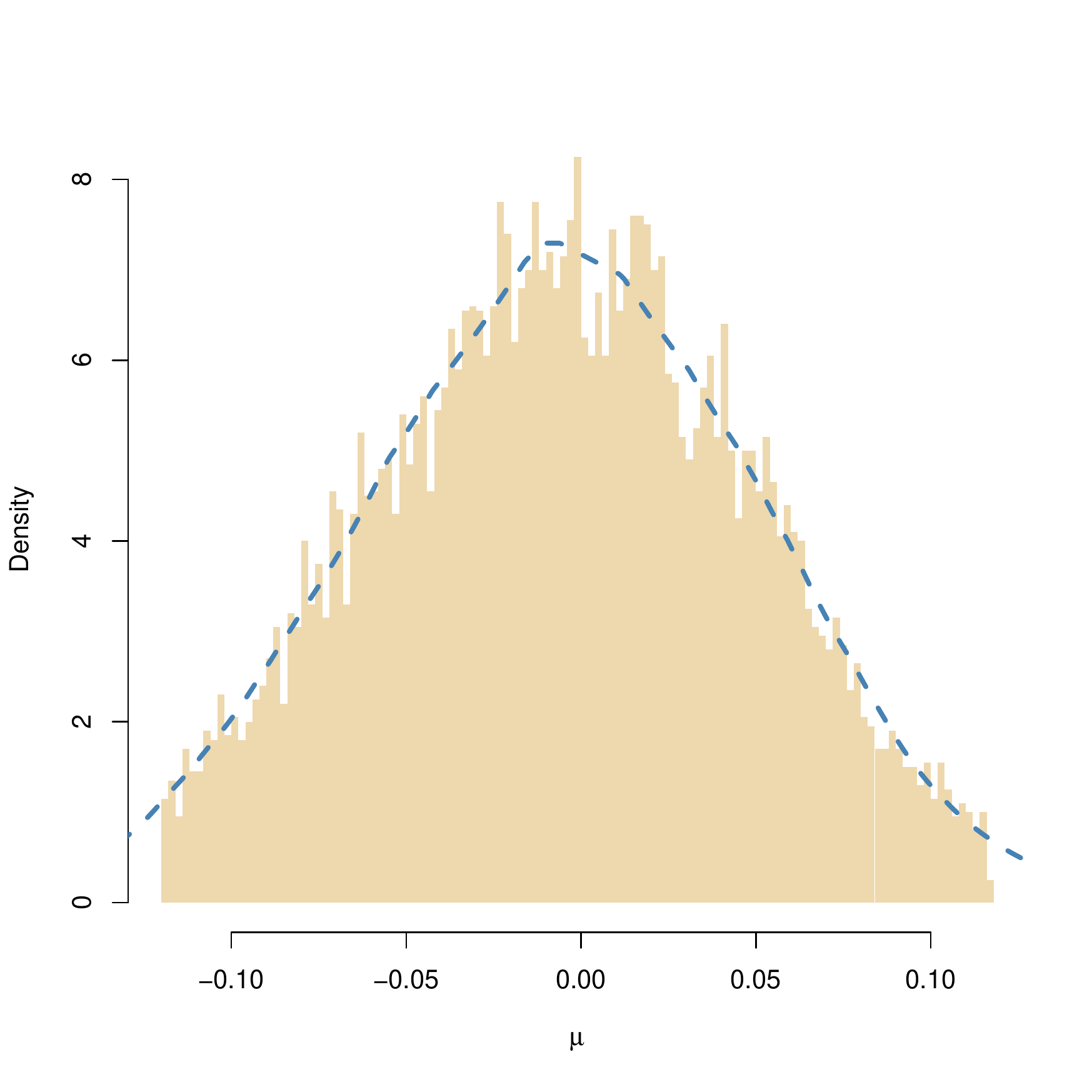}}


While this bridge solution produces valuable approximations when both
parameters $\theta_0$ and $\theta_1$ are within the same parameter space and
have the same or similar absolute meanings (e.g., $\theta$ is equal to
$\mathbb{E}_\theta[X]$ in both models), it does not readily apply to settings
with variable dimension parameters.  In such cases, separate approximations of
the evidences, i.e.~of the numerator and denominator in $\mathfrak{B}_{01}$ are
requested, with the exception of reversible jump Monte Carlo techniques
\citep{green:1995} presented in the following section. Although using harmonic
means for this purpose as in \citet{newton:raftery:1994} is fraught with
danger, as discussed in \citet{neal:1994} and \citet{marin:robert:2010}, we
refer the reader to this later paper of ours for a model-based solution using
an importance function restricted to an HPD region (see also
\citealp{robert:wraith:2009} and \citealp{weinberg:2012}).  We however insist
on (and bemoan) the lack of generic solution for the approximation of Bayes
factors, despite those being the workhorse of Bayesian model selection and
hypothesis testing.

\section{MCMC methodology}\label{sec:MCMC}

The above Monte Carlo techniques impose (or seem to impose) constraints on the
posterior distributions that can be approximated by simulation. Indeed, direct
simulation from this target distribution is not always feasible in a (time-wise) manageable
form, while importance sampling may result in very poor or even worthless approximations,
as for instance when the empirical average
$$
\dfrac{1}{T} \sum_{t=1}^T \dfrac{f(x_t)}{g(x_t)} h(x_t)
$$
suffers from an infinite variance. Finding a reliable importance function
thus requires some sufficient knowledge about the posterior density
$\pi(\cdot|x)$. Markov chain Monte Carlo (MCMC) methods
were introduced (also in Los Alamos) with the purpose of bypassing this requirement
of an a priori knowledge on the target distribution. On principle, they apply
to any setting where $\pi(\cdot|x)$ is known up to a normalising constant (or
worse, as a marginal of a distribution on an augmented space). 

As described in another chapter of this volume (Craiu and Rosenthal, 2013), MCMC methods rely on
ergodic theorems, i.e.~the facts that, for positive recurrent Markov chains,
(a) the limiting distribution of the chain is always the stationary
distribution and (b) the law of large numbers applies. The fascinating feature
of those algorithms is that it is straightforward to build a Markov chain
(kernel) with a stationary distribution equal to the posterior distribution,
even when the latter is only know up to a normalising constant. Obviously,
there are caveats with this rosy tale: complex posteriors remain harder to
approximate than essentially Gaussian posteriors, convergence (ergodicity) may
require in-human time ranges or simply not agree with the limited precision of
computers.

For completeness' sake, we recall here the format of a random walk Metropolis--Hastings 
(RWMH) algorithm \citep{hastings:1970}

\begin{algorithm}
\caption{RWMH}
\begin{algorithmic}
\FOR {$t=1$ to $T$}
\STATE Generate $\xi \sim \varphi(|\xi-\theta_{t-1}|)$
\STATE Take $\theta_t=\xi$ with probability $\alpha=\min\{1,
f_0(\bx|\xi)\pi_0(\xi) \big/ f_0(\bx|\theta_{t-1})\pi_0(\theta_{t-1})$
\STATE Take $\theta_t=\theta_{t-1}$ otherwise.
\ENDFOR
\end{algorithmic}
\end{algorithm}

\piccaption{\label{fig:RWMHL}
{\small Values of the Markov chain $(\mu_{t})$ {\em (sienna)} and of iid simulations {\em (wheat)}
for $10^3$ iterations and a double exponential sample of size $n=150$, when using a RWMH
algorithm with scale equal to $1$.}}
\piccaptioninside
\parpic[r]{\includegraphics[width=5cm]{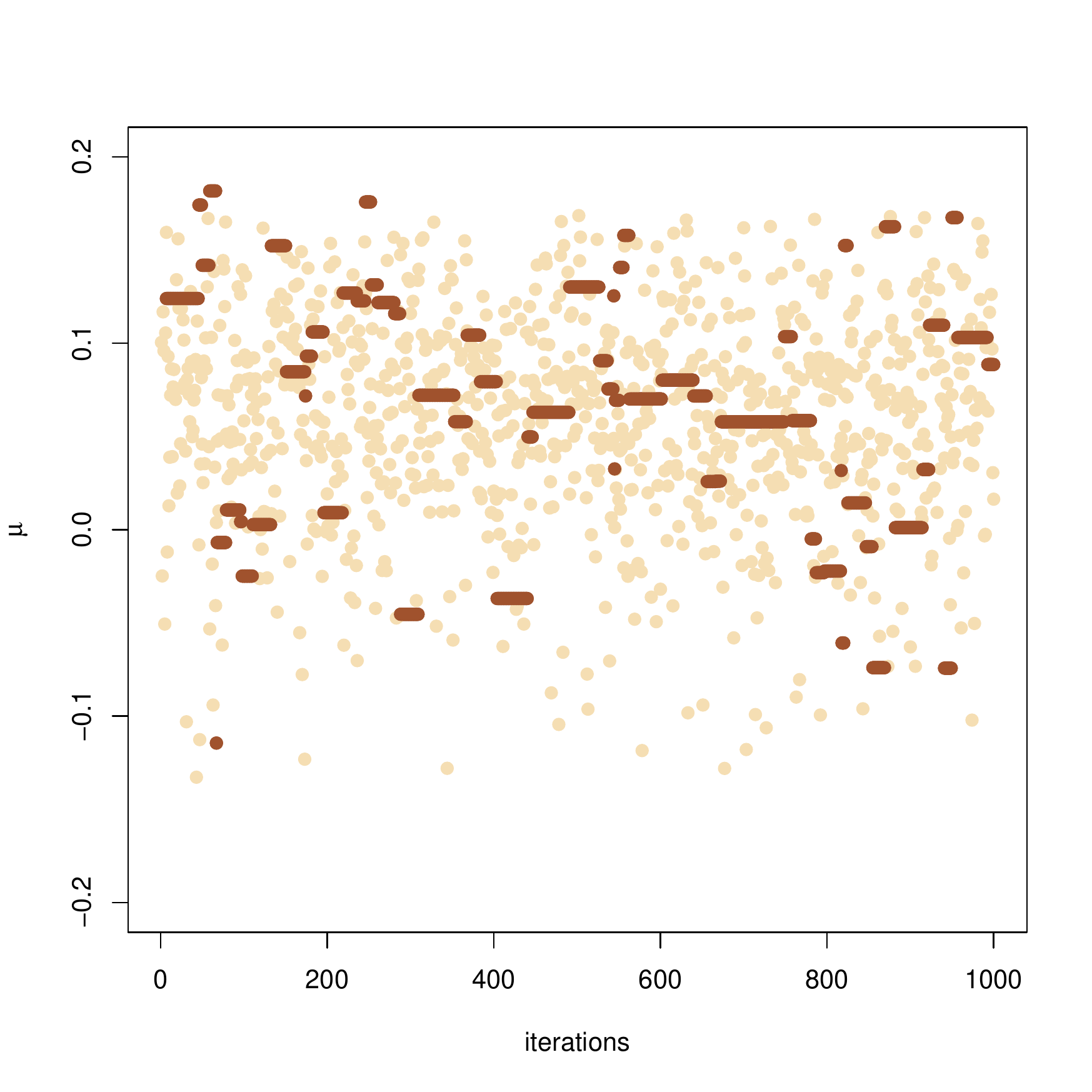}}

\begin{example}[Example 1 (quater)] If we consider once again the posterior distribution
on $\mu$ associated with a Laplace sample, even though the exact simulation from this
distribution was implemented in Example 1 (ter), an MCMC implementation is readily available.
Using a RWMH algorithm, with a normal distribution centred at $\mu_{t-1}$
and with scale $\sigma$, the implementation of the method is straightforward. 

As shown on Figure \ref{fig:RWMHL}, the algorithm is less efficient than an iid sampler, with an acceptance
rate of only $6\%$. However, one must also realise that devising the code behind the algorithm
only took five lines and a few minutes, compared with the most elaborate construction behind the
iid simulation!
\end{example}


\subsection{Gibbs sampling}
A special class of MCMC methods seems to have been especially designed for Bayesian
hierarchical modelling (even though they do apply in a much wider generality). Those
go under the denomination of Gibbs samplers, unfortunately named after Gibbs for
the mundane reason that one of their initial implementations was for the simulation of
Gibbs random fields (in image analysis, \citealp{geman:geman:1984}). Indeed, Gibbs
sampling addresses the case of (often) high-dimensional problems found in hierarchical 
models where each parameter (or group of parameters) is endowed with a manageable full
conditional posterior distribution. (While the joint posterior is not manageable.)
The principle of the Gibbs sampler is then to proceed by local simulations from those
full conditionals in a rather arbitrary order, producing a Markov chain whose stationary
distribution is the joint posterior distribution.

Let us recall that a Bayesian hierarchical model is build around a hierarchy of probabilistic
dependences, each level depending only on the neighbourhood levels (except for global parameters
that may impact all levels). For instance,
$$
\bx \sim f(\bx|\theta_1)\,,\ \theta_1|\theta_2 \sim \pi_1(\theta_1|\theta_2)\,,\
\theta_2\sim\pi_2(\theta_2)
$$
induces a simple hierarchical model in that $\bx$ only depends on $\theta_1$ while $\theta_2$
only depends on $\theta_1$---i.e., $\bx$ is independent of $\theta_2$ given $\theta_1$.

Examples of such structures abound: 

\begin{example}[Example 2]
A typical instance is made of random effect models as in the
following instance (inspired from \citealp{breslow:clayton:1993}) of Poisson observations
$(i=1,\ldots,n,\,j=1,\ldots,N_j)$
\begin{align*}
x_{ij} 	      &\sim \mathcal{P}(\exp\{\mu_i+\epsilon_{ij}\})\\
\epsilon_{ij} &\sim \mathcal{N}(0,\rho^2)\\
\mu_i &= \log m_i + \bz_i^\text{T}\beta\\
\beta &\sim \mathcal{N}_d(0,\sigma^2 \mathbf{I}_d)\\
\sigma^2,\rho^2 &\sim \pi(\omega) = 1/\omega 
\end{align*}
where $i$ denotes a group or district label, $j$ the replication index, $\bz_i$ a vector
of covariates, $m_i$ a population size. In this model, given the data $\bx=\{x_{ij}, 
i=1,\ldots,n,\,j=1,\ldots,N_j\}$, a Gibbs sampler generates from the joint distribution 
of $\epsilon_{ij}$, $\beta$, $\sigma^2$, and $\rho^2$ by using the conditionals
\begin{align*}
\epsilon_{ij} &\sim \pi(\epsilon_{ij}|x_{ij},\mu_i,\rho^2)\\
\beta &\sim \pi(\beta|\bx,\beps,\sigma^2)\\
\rho^2 &\sim \pi(\rho^2|\beps)\\
\sigma^2 &\sim \pi(\sigma^2|\beta)
\end{align*}
which are more or less manageable (as they may require individual Metropolis--Hasting implementations where
the Poisson distribution is replaced with its normal approximation in the proposal). Note, however, that this
simple solution hides a potential difficulty with the choice of an improper prior on $\sigma^2$ and $\rho^2$.
Indeed, even though the above conditionals are well-defined for all samples, it may still be that the associated
joint posterior distribution does not exist. This phenomenon of the {\em improper posterior} was exhibited in
\citet{casella:george:1992} and analysed in \citet{hobert:casella:1996}.
\end{example}

\begin{figure}
\centerline{\includegraphics[width=10cm]{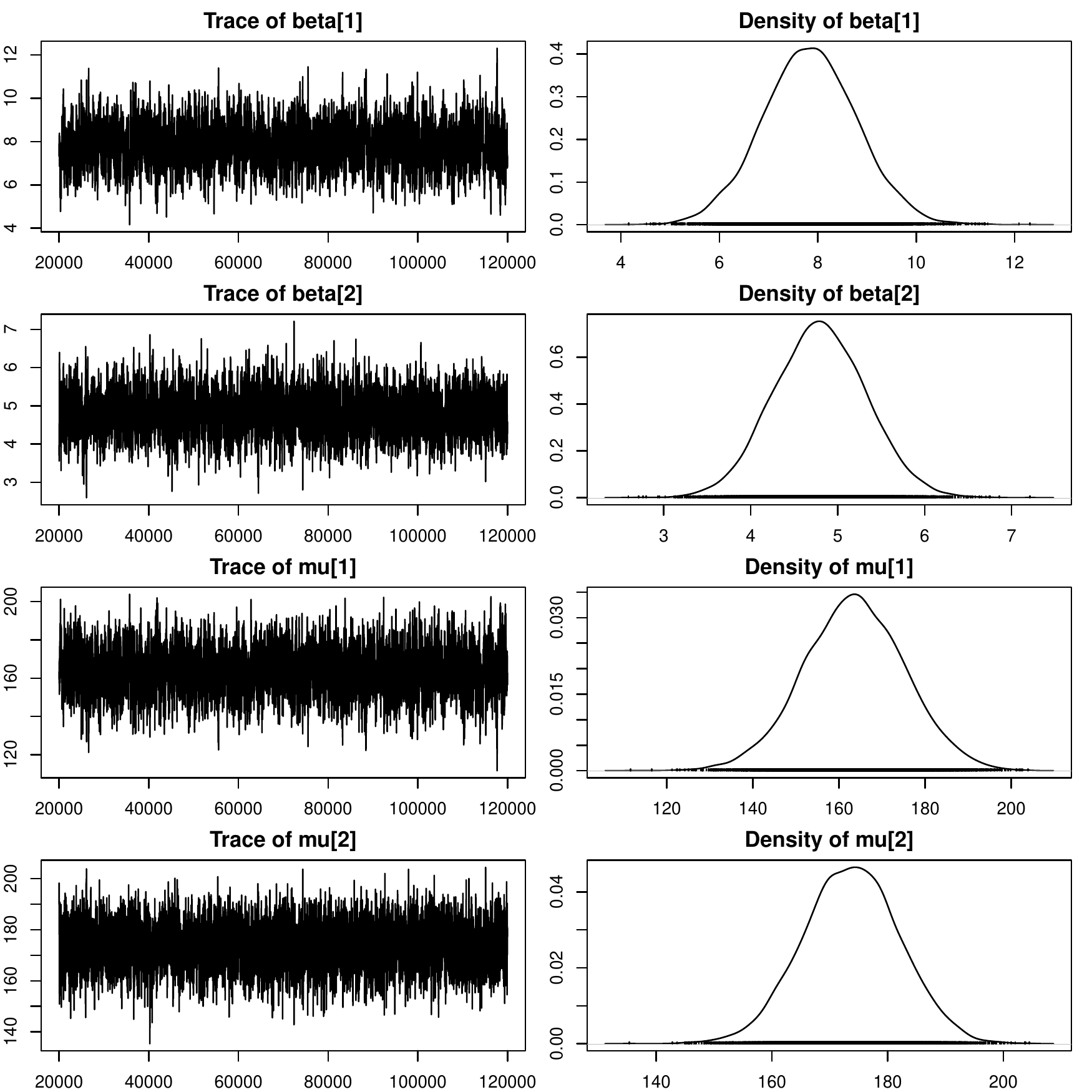}}
\caption{\label{fig:pof1}
{\small Evolution of the Gibbs Markov chains for some parameters of the growth mixed-effect
model of Pothoff and Roy (1964) {\em (right)} and density estimate of the corresponding posterior distribution {\em (right)},
based on $120,000$ iterations.}}
\end{figure}

\begin{figure}
\centerline{\includegraphics[width=10cm]{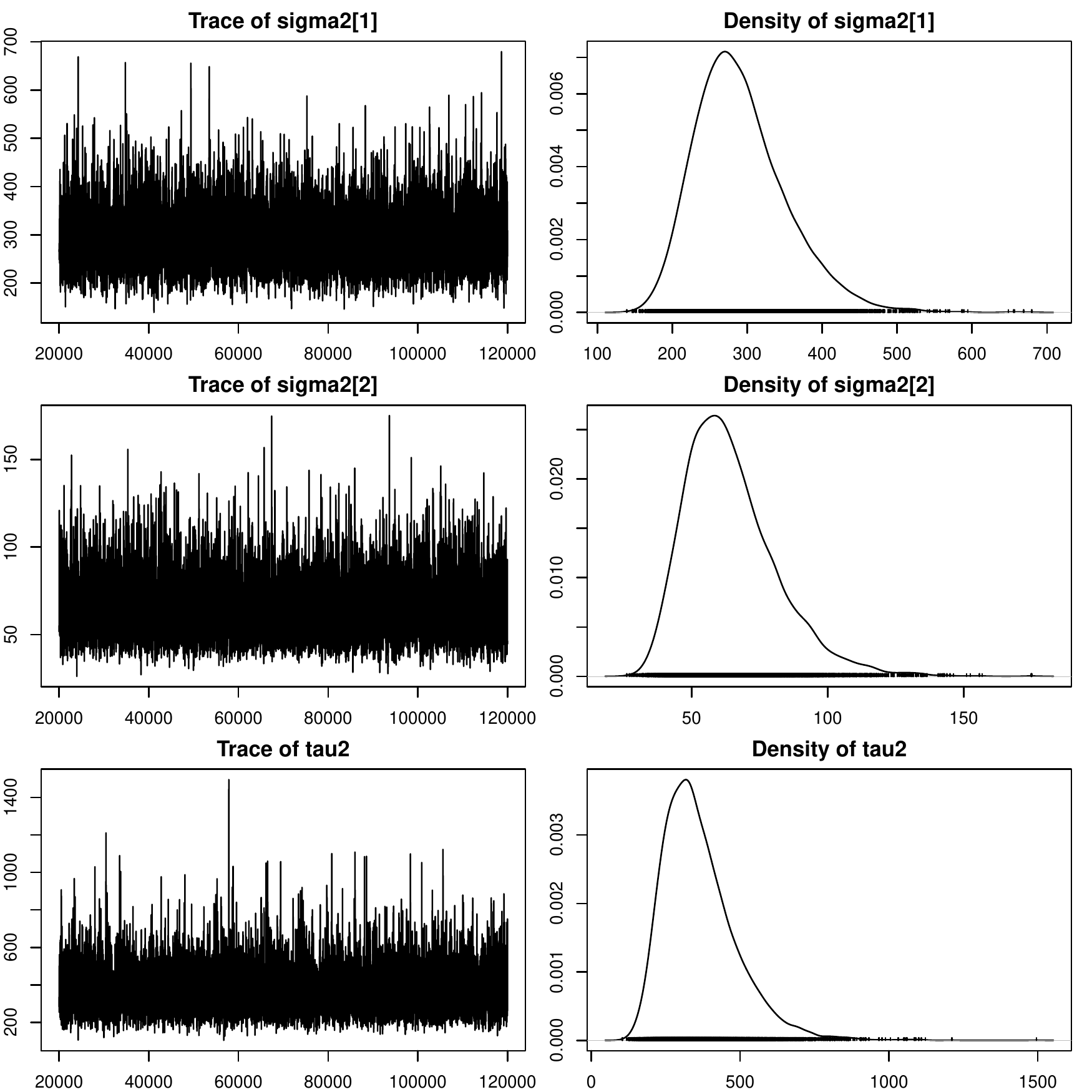}}
\caption{\label{fig:pof2}
{\small Same legend as Figure \ref{fig:pof1}.}}
\end{figure}

\begin{example}[Example 3]
A growth measurement model was applied by \citet{pothoff:roy:1964} to dental measurements
of 11 girls and 16 boys, as a mixed-effect model. (The dataset is available in {\sf R} as
{\sf orthodont} in package {\sf nlme}.) Compared with the random effect models,
mixed-effect models include additional random-effect terms and are more
appropriate for representing clustered, and therefore dependent, data arising in, e.g.,
hierarchical, paired, or longitudinal data.) For $i=1,\ldots,n$ children and $j=1,\ldots,r$ 
observations on each child, growth is expressed as
\piccaption{\label{fig:dagoff}
{\small Directed acyclic graph associated with the Bayesian modelling of the growth data of
\citet{pothoff:roy:1964}.}}
\piccaptioninside
\parpic[r]{\includegraphics[width=7cm]{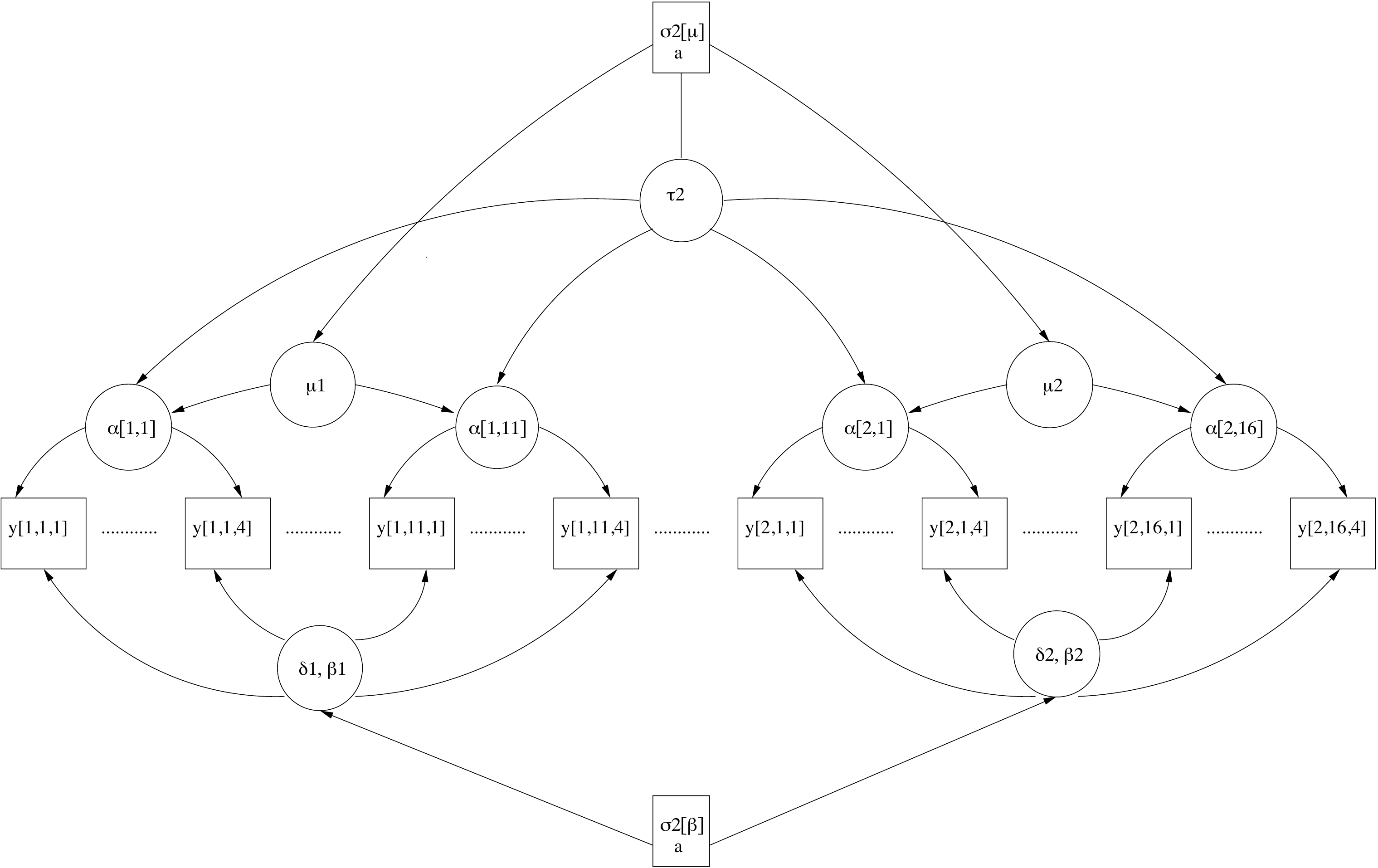}}
\newcommand\bh{\mathbf{h}}
\newcommand\bt{\mathbf{t}}
$$
y_{ij}=\alpha_i+\beta_{h_i}t_j+\sigma^2_{h_i}\epsilon_{ij}\,,
$$
where $\bh=(h_1,\ldots,h_n)$ is a sex factor with $h_i\in\{1,2\}$
($1$ corresponds to female and $2$ to male) and $\bt=(t_1,\ldots,t_r$ is the vector of ages.
The random effects in this growth model are the $\alpha_i$'s, which are independent
$\mathcal{N}\left(\mu_{h_i},\tau^2\right)$ variables. The priors on the corresponding parameters
are chosen to be conjugate:
$$
\beta_1,\beta_2\sim\mathcal{N}_1\left(0,\sigma^2_\beta\right)\,,\quad 
\sigma^2_1,\sigma_2^2,\tau^2\sim\mathcal{IG}(a,a)\,,\quad \sigma^2_2\sim\mathcal{IG}(a,a)\,,
\quad
\mu_1,\mu_2\sim\mathcal{N}_1\left(0,\sigma^2_\mu\right)\,,
$$
where $\mathcal{IG}(a,a)$ denotes the inverse gamma distribution. Note that,
while the posterior distribution is well-defined in this case, there is no
garantee that the limit exists when $a$ goes to zero and thus that small values
of $a$ should be avoided as they do not necessarily constitute proper default
values.  Figure \ref{fig:dagoff} summarises the Bayesian model through a DAG
(directed acyclic graph, see \citep{lauritzen:1996}).

Thanks to this conjugacy, the full conditionals are available as standard distributions
$(k=1,2)$:
\begin{align*}
\beta_k &\sim \mathcal{N}\left(\dfrac{\sum_{j=1}^r t_j
\sum_{i=1}^n\mathbb{I}_{h_i=k}(y_{ij}-\alpha_i)\sigma^{-2}_1}
{n_k\sum_{j=1}^r t_j^2\sigma^{-2}_1+\sigma^{-2}_\beta},
\left\{{n_k\sum_{j=1}^r t_j^2}{\sigma^{-2}_1}+{\sigma^{-2}_\beta}\right\}^{-1}\right)\\
\sigma^2_k &\sim \mathcal{IG}\left(a+\nicefrac{n_kr}{2},a+\sum_{i=1}^n\mathbb{I}_{h_i=k}
\sum_{j=1}^r\left(y_{ij}-\beta_1t_j-\alpha_i\right)^2\big/2\right)\\
\mu_k &\sim \mathcal{N}\left(\dfrac{\left(\sum_{i=1}^n\mathbb{I}_{h_i=k}\alpha_i\right)\tau^{-2}}
{n_k\tau^{-2}+\sigma^{-2}_\mu},\left\{{n_k}{\tau^{-2}}+{\sigma^{-2}_\mu}\right\}^{-1}\right)\\
\tau^2 &\sim \mathcal{IG}\left(a+\nicefrac{n}{2},a+\sum_{i=1}^n(\alpha_i-\mu_{h_i})^2\big/2\right)\,,
\end{align*}
where $n_k$ is the number of children with sex $k$, and $(i=1,\ldots,n)$
$$
\alpha_i \sim \mathcal{N}\left(\frac{\sum_{j=1}^r
(y_{ij}-\beta_{h_i}t_j)\sigma^{-2}_{h_i}+\mu_{h_i}\tau^{-2}}{\tau^{-2}+r\sigma^{-2}_{h_i}},
\left\{\tau^{-2}+r\sigma^{-2}_{h_i}\right)^{-1}\right\}\,.
$$
It is therefore straightforward to run the associated Gibbs sampler. Figures \ref{fig:pof1} and
\ref{fig:pof2} show the raw output of some parameter series, based on $120,000$ iterations.
For instance, those figures show that $\beta_1$ and $\beta_2$ are possibly equal, as their likely
ranges overlap. This does not seem to hold for $\mu_1$ and $\mu_2$.
\end{example}

One of the obvious applications of the Gibbs sampler is found in graphical models---an
application that occurred in the early days of MCMC---since those models are defined by and
understood via conditional distributions rather than through an unmanageable joint distribution.
As detailed in \citet{lauritzen:1996}, undirected probabilistic graphs are Markov with respect to
the graph structure, which means that variables indexed by a given node $\eta$
of the graph only depend on variables indexed by nodes connected to $\eta$. For instance, if the
vector indexed by the graph is Gaussian, $\bX\sim\mathcal{N}(\mu,\Upsigma)$, the non-zero terms of
$\Upsigma^{-1}$ correspond to the edges of the graph. Applications of this modelling abound, as for
instance in experts systems \citep{spiegelhalter:dawid:lauritzen:cowell:1993}. Note that hierarchical
Bayes models can be naturally associated with dependence graphs leading to DAGs and thus fall within this category
as well.

\subsection{Reversible-jump MCMC}\label{sub:rjMCMC}

Although the principles of the MCMC methodology are rather straightforward to
understand and to implement, resorting for instance to down-the-shelf
techniques like RWMH algorithms, a more challenging setting occurs with the
case of variable dimensional problems. These problems typically occur in a
Bayesian model choice situation, where several (or an infinity of) models are
considered at once. The resulting parameter space is a {\em millefeuille}
collection of sets, with most likely different dimensions, and moving around
this space or across those layers is almost inevitably a computational issue.
Indeed, the only case open to direct computation is the one when the posterior
probabilities of the models under comparison can be evaluated, resulting in a
two-stage implementation, the model being chosen first and the parameters of
this model being simulated ``as usual". However, as seen above, computing
posterior probabilities of models is rarely a straightforward case. In other
settings, moving around the collection of models and within the corresponding
parameter spaces must occur simultaneously, especially when the number of
models is large or infinite.

Defining a Markov chain kernel that explores the multi-layered space is challenging
because of the difficulty of defining a reference measure on this complex space. However,
\citet{green:1995} came up with a solution that is rather simplex to express (if not necessarily
to implement). The idea behind Green's (1995) reversible jump solution is to take advantage of
the Markovian nature of the algorithm: since all that matters in a Markov chain is the most
recent value of the chain, exploration of a multi-layered space, represented as a direct sum \citep{rudin:1976}
of those spaces,
$$
\bigoplus_{i=1}^I \Uptheta_i\,,
$$ 
only involves a pair of sets $\Uptheta_i$ at each step, $\Uptheta_\upiota$ and $\Uptheta_\uptau$ say.
\newcommand{\uio}{\upiota}
\newcommand{\uto}{\uptau}
Therefore, the mathematical difficulty reduces to create a connection between both spaces, difficulty
that is solved by Green's (1995) via the introduction of auxiliary variables $\lambda_\uio$ and
$\lambda_\uto$ in order for $(\theta_\uio,\lambda_\uio)$ and $(\theta_\uto,\lambda_\uto)$ to be
in one-to-one correspondence, i.e.~$(\theta_\uio,\lambda_\uio) = \Psi(\theta_\uto,\lambda_\uto)$.
Arbitrary distributions on $\lambda_\uio$ and on $\lambda_\uto$ then come to complement the
target distributions $\pi(\uio,\theta_\uio|x)$ and $\pi(\uto,\theta_\uto|x)$. The algorithm is call
reversible because the symmetric move from $(\theta_\uio,\lambda_\uio)$ to $(\theta_\uto,\lambda_\uto)$
must follow $(\theta_\uto,\lambda_\uto) = \Psi^{-1}(\theta_\uio,\lambda_\uio)$. In other words, moves one way
determine moves the other way. A schematic representation is as follows:

\begin{algorithm}
\caption{RJMCM}
\begin{algorithmic}
\FOR {$t=1$ to $T$}
\STATE Given current state $(\uio,\theta_\uio)$,
\STATE Generate index $\uto$ from the prior probabilities $\pi(\uto)$.
\STATE Generate $\lambda_\uio$ from the auxiliary distribution $\pi_\uio(\lambda_\uio)$
\STATE Compute $(\theta_\uto,\lambda_\uto) = \Psi^{-1}(\theta_\uio,\lambda_\uio)$
\STATE Accept to switch to $(\uio,\theta_\uio)$ with probability 
$$
\alpha = \dfrac{\pi(\uto,\theta_\uto|x) \pi_\uto(\lambda_\uto)}{\pi(\uio,\theta_\uio|x) \pi_\uio(\lambda_\uio)}\,
\left| \dfrac{\text{d}\Psi(\theta_\uto,\lambda_\uto)}{\text{d}(\theta_\uto,\lambda_\uto)} \right|
$$
\STATE Else reproduce $(\uio,\theta_\uio)$
\ENDFOR
\end{algorithmic}
\end{algorithm}

The important feature in the above acceptance probability is the Jacobian 
term $\text{d}\Psi(\theta_\uto,\lambda_\uto)\big/\text{d}(\theta_\uto,\lambda_\uto)$
which corresponds to the change of density in the transformation. It is also a source
of potential mistakes in the implementation of the algorithm.

The simplest version of RJMCM is when $\theta_\uto = (\theta_\uio,\lambda_\uio)$, i.e.~when
the move from one parameter space to the next involves adding or removing one parameter, as
for instance in estimating a mixture with an unknown number of components \citep{richardson:green:1997}
or a $MA(p)$ time series with $p$ unknown. It can also be used with $p$ known, as illustrated below.

\begin{example}[Example 4] 
An $MA(p)$ time series model---where MA stands for `moving average'---is defined by the equations
$$
x_t = \sum_{i=1}^p \vartheta_i \epsilon_{t-i} + \epsilon_t
\qquad t=1,\ldots\,,
$$
where the $\epsilon_t$'s are iid $\mathcal{N}(0,\sigma^2)$.
While this model can be processed without RJMCMC, we present here a resolution explained in
\citet{marin:robert:2007} that does not distinguish between the cases when $p$ is known and when
$p$ is unknown. 

\newcommand{\bB}{\mathbf{B}}
\newcommand{\bI}{\mathbf{I}}
The associated ``lag polynomial" $\mathcal{P}(\bB) = \bI + \sum_{i=1}^p \vartheta_i \bB^i$ provides a formal
representation of the series as $x_t=\mathcal{P}(\bB)\epsilon_t$, with $\bI\epsilon_t=\epsilon_{t}$,
$\bB\epsilon_t=\epsilon_{t-1}$, ... As
a polynomial it also factorises through its roots $\uplambda_i$ as 
$$
\mathcal{P}(\bB) = \prod_{i=1}^p (\bI-\uplambda_i\bB)\,.
$$
While the number of roots is always $p$, the number of (non-conjugate) complex
roots varies between $0$ (meaning no complex root) and $\lfloor \nicefrac{p}{2}
\rfloor$. This representation of the	model thus induces a variable dimension
structure in that the parameter space is then the product $(-1,1)^r\times
B(0,1)^{\nicefrac{p-r}{2}}$, where $B(0,1)$ denotes the complex unit ball and
$r$ is the number of real valued roots $\uplambda_iB$. The prior distributions
on the real and (non-conjugate) complex roots are the uniform distributions on
$(-1,1)$ and $B(0,1)$, respectively.  In other words, 
\begin{equation}\label{eq:AprioR} 
\pi(\boldsymbol{\lambda}) =
\frac{1}{\lfloor \nicefrac{p}{2} \rfloor+1}\, \prod_{\uplambda_i\in{(-1,1)}} \frac{1}{2}{\mathbb I}_{|\uplambda_i|<1}
\,\prod_{\uplambda_i\not\in{\mathbb R}} \frac{1}{\pi}{\mathbb I}_{B(0,1)}(\uplambda_i)\,, 
\end{equation} 
Moving around this space using RJMCMC is rather straightforward: either the number of real roots does not change
in which case any regular MCMC step is acceptable or the number of real roots moves up or down by a factor of 2,
new roots being generated from the prior distribution, in which case the above RJMCMC acceptance ratio reduces to
a likelihood ratio. An extra difficulty with the $MA(p)$ setup is that the likelihood is not available in closed
form unless the past innovations $\epsilon_{0},\epsilon_{-1},\ldots,\epsilon_{1-p}$ are available. As explained 
in \citet{marin:robert:2007}, they need to be simulated in a Gibbs step, that is, conditional upon the other parameters
with density proportional to
$$
\prod_{t=0}^{1-p}\exp\left\{-\nicefrac{\epsilon_{t}^2 }{ 2\sigma^2 } \right\} \,
\prod_{t=1}^T\exp\left\{-\left(x_t-\mu+ \sum_{j=1}^p \vartheta_j\widehat \epsilon_{t-j}
                \right)^2\bigg/ 2\sigma^2 \right\} \,,
$$
where $\hat \epsilon_{0}=\epsilon_0$,$\ldots$, $\hat \epsilon_{1-p}=\epsilon_{1-p}$ and $(t>0)$
$$
\widehat \epsilon_t = x_t -\mu + \sum_{j=1}^p \vartheta_j \widehat\epsilon_{t-j}\,.
$$
This recursive definition of the likelihood is rather costly since it involves computing the 
$\widehat \epsilon_t$'s for each new value of the past innovations, hence $T$ sums of $p$ terms.
Nonetheless, the complexity $\text{O}(Tp)$ of this representation is much more manageable than the normal 
exact representation mentioned above.
\end{example}

As mentioned above, the difficulty with RJMCM is in moving from the general
principle (which indeed allows for a generic exploration of varying dimension
spaces) to the practical implementation: when faced with a wide range of
models, one needs to determine which models to pair together---they must be
similar enough---and how to pair them---so that the jumps are efficient enough.
This requires the calibration of a large number of proposals, whose efficiency
is usually much lower than in single-model implementations. Whenever the number
of models is limited, my personal experience is that it is more efficient to
run separate (and parallel) MCMC algorithms on all models and to determine the
corresponding posterior probabilities of those models by a separate evaluation,
like Chib's (\citeyear{chib:1995}). (Indeed, a byproduct of the RJMCMC
algorithm is to provide an evaluation of the posterior probabilities of the
models under comparison via the frequencies of accepted moves to such models.)
See, e.g., \citet{lee:marin:mengersen:robert:2008} for an illustration in the
setting of mixtures of distributions. We end up with a word of caution against
the misuse of probabilistic structures over those collections of spaces, as illustrated
by \cite{congdon:2006} and \cite{scott:2002} \citep{robert:marin:2008}.


\section{Approximate Bayesian computation methods}\label{sec:ABC}

This section covers some aspects of a specific computational method called
Approximate Bayesian computation (ABC in short), which stemmed from acute
computational problems in statistical population genetics and rised in
importance over the past decade. The section should be more methodological than
the previous sections as the method is not covered in this volume, as far as I
can assess. In addition, this is a special computational method in that it has
been specifically developed for challenging Bayesian computational problems
(and that it carries the Bayesian label within its name!). Although the reader
is referred to, e.g., \citet{toni:etal:2009} and \citet{beaumont:2010} for a
deeper review on this method, I will cover here different accelerating
techniques and the numerous calibration issues of selecting both the tolerance
and the summary statistics. 

Approximate Bayesian computation (ABC) techniques appeared at the end of the
20th Century in population genetics
\citep{tavare:balding:griffith:donnelly:1997,pritchard:seielstad:perez:feldman:1999},
where scientists were faced with intractable likelihoods that MCMC methods were
simply unable to handle with the slightest amount of success. Some of those
scientists developed simulation tools overcoming the jamming block of
computing the likelihood function that turned into a much more general form of
approximation technique, exhibiting fundamental links with econometric methods
such as indirect inference \citep{gourieroux:monfort:renault:1993}. Although some part of the statistical community was
initially reluctant to welcome them, trusting instead massively parallelised
MCMC approaches, ABC techniques are now starting to be part of the statistical
toolbox and to be accepted as an inference method {\em per se}, rather than
being a poor man's alternative to more mainstream techniques. While details
about the method are provided in recent surveys
\citep{beaumont:2008,beaumont:2010,marin:pudlo:robert:ryder:2011}, we expose in
algorithmic terms the basics of the ABC algorithm:

\begin{algorithm}
\caption{ABC}
\begin{algorithmic}
\FOR {$t=1$ to $T$}
\REPEAT
\STATE Generate $\theta^{*}$ from the prior $\pi(\cdot)$.
\STATE Generate $x^{*}$ from the model $f(\cdot|\theta^{*})$.
\STATE Compute the distance $\rho(S(\mathbf{x}^0),S(\mathbf{x}^{*}))$.
\STATE Accept $\theta^{*}$ if $\rho(S(\mathbf{x}^0),S(\mathbf{x}^{*}))<\epsilon$.
\UNTIL acceptance
\ENDFOR
\end{algorithmic}
\end{algorithm}

The idea at the core of the ABC method is to replace an acceptance based on the unavailable likelihood with
one evaluating the pertinence of the parameter from the proximity between the data and a simulated pseudo-data.
This proximity is using a distance or pseudo-distance $\rho(\cdot,\cdot)$ between a (summary) statistic $S(x^0)$
based on the data and its equivalent $S(x{*}$ for the pseudo-data. We stress from this early stage that the summary
statistic $S$ is very rarely sufficient and hence that ABC looses some of the information contained in the data.

\begin{example}[Example 4 (bis)]
While the MA$(p)$ is manageable by other approaches---since the missing data
structure is of a moderate complexity---, it provides an illustration of a
model where the likelihood function is not available in closed form and where
the data can be simulated in a few lines of code given the parameter. Using the
$p$ first autocorrelations as summary statistics $S(\cdot)$, we can then
simulate parameters from the prior distribution and corresponding series
$\bx^*=(x_1^*,\ldots,x^*_T)$ and only keep the parameter values associated with
the smallest $S(\bx^*)$'s. 

As shown in Figure \ref{fig:rawdist}, reproduced from
\citet{marin:pudlo:robert:ryder:2011}, there is a difference between the genuine
posterior distribution and the ABC approximation, whatever the value of $\epsilon$
is. This comparison also shows that the approximation stabilises quite rapidly 
as $\epsilon$ decreases to zero, in agreement with the general argument that the
tolerance should not be too close to zero for a given sample size \citep{fearnhead:prangle:2012}.
\end{example}
\piccaption{\label{fig:rawdist}
{\small Variation of the estimated distributions of ABC samples using different
quantiles on the simulated distances for $\epsilon$ ($10\%$ \textit{in blue},
$1\%$ \textit{in red}, and $0.1\%$ \textit{in yellow}) when compared with the
true marginal densities. The observed dataset is simulated from an MA$(2)$
model with $n = 100$ observations and parameter $\vartheta = (0.6, 0.2)$ 
{\em (Source: \citealp{marin:pudlo:robert:ryder:2011}).}}}
\piccaptioninside
\parpic[r]{\includegraphics[width=7cm]{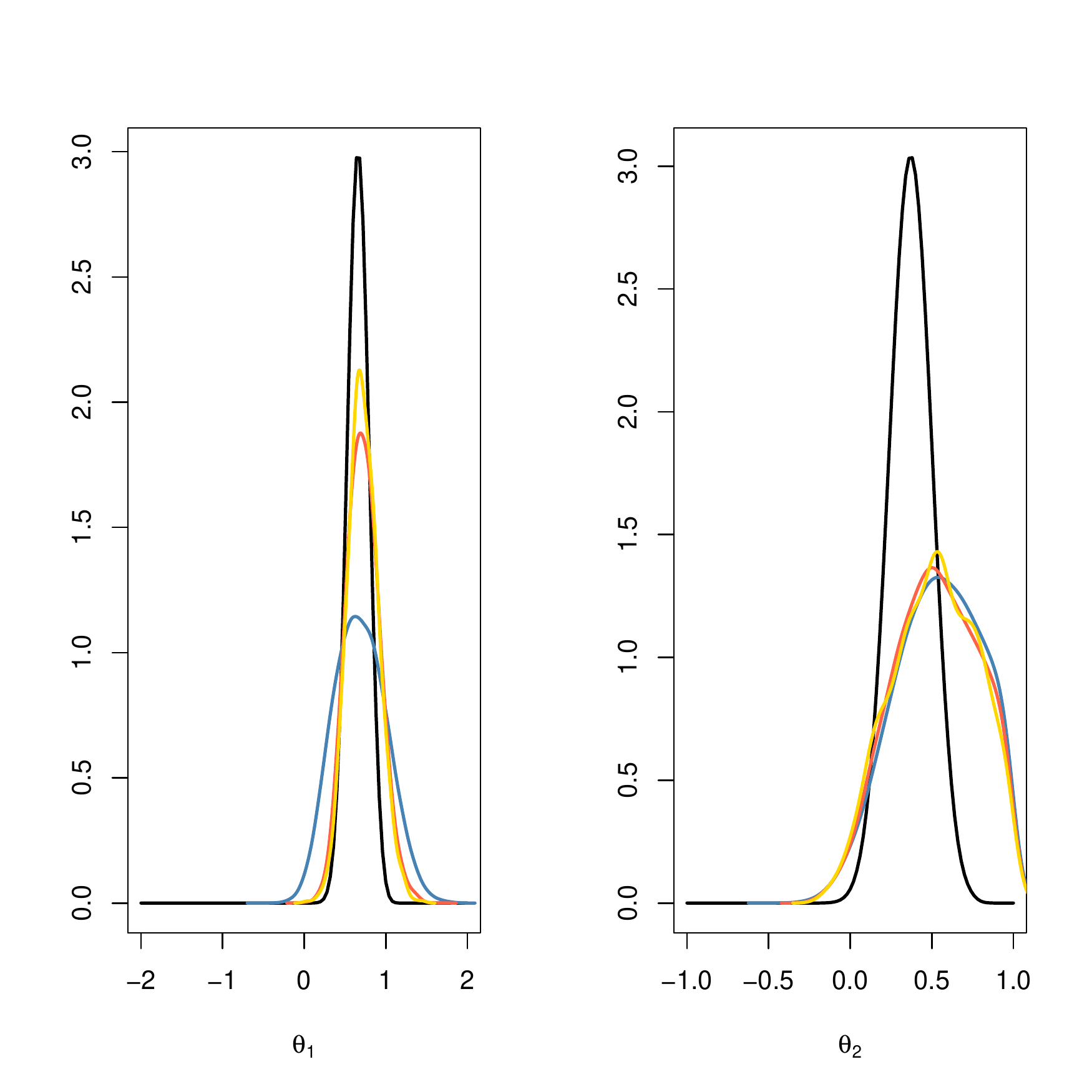}}

ABC suffers from an ``information paradox", namely that it quickly stops to pay to
increase the dimension of the summary statistic $S(\cdot)$ in the hope to
bring the ABC inference closer to a ``perfect" Bayesian inference based on the
whole data and thus fill the information gap. For one thing,
increasing the dimension of the summary statistic
invariably leads to increase the tolerance $\epsilon$, as discussed
below.

For another thing, considering the most extreme case illuminates this paradox.
As noted above, ABC is almost always based on summary statistics, $S(\cdot)$,
rather than on the raw data. The reason why is obvious in Example 4 (bis),
since using the raw time series instead of the vector of empirical
autocorrelations would have been strongly detrimental as the distance between
two simulated series grows with the time horizon and brings very little
information about the value of the underlying parameter. In other words, it
forces us to use a much larger tolerance $\epsilon$ in the algorithm. The
paradox is easily explained by the following points:

\begin{itemize}
\item[--] the (initial) intuition upon which ABC is built considers the limiting case $\epsilon\approx 0$ and
the fact that $\piabc(\cdot|\bx^0)$ is an approximation to $\pi(\cdot|\bx^0)$, as opposed to the true setting being
that $\piabc(\cdot|S(\by))$ is an approximation to $\pi(\cdot|S(\bx^0))$ and that it incorporates a Monte Carlo 
error as well;
\item[--] for a given computational effort, the tolerance $\epsilon$ is necessarily positive---if only to produce
a positive acceptance rate---and deeper studies show that it behaves like a
non-parametric bandwidth parameter, hence increasing with the dimension of $S$ while (slowly) decreasing 
with the sample size. 
\end{itemize}

Therefore, when the dimension of the raw data is large (as for instance in the time series setting
of Example 4 bis), it is definitely not recommended
to use a distance between the raw data $\bx^0$ and the raw pseudo-data $\bx^*$: the {\em curse of
dimension}\index{curse of dimension} operates in nonparametric statistics and clearly impacts the approximation
of $\pi(\cdot|\bx^0)$ as to make it impossible even for moderate dimensions. 

In connection with the above, it must be stressed that, in almost
any implementation, the ABC algorithm is not {\em correct} for at least two reasons: the data $\bx^0$
is replaced with a roughened version $\{\bx^*; s\rho(S(\bx^0),S(\bx^*))<\epsilon\}$ and the use of a non-sufficient
summary statistic $S(\cdots)$. In addition, as in regular Monte Carlo approximations, a given computational effort 
produces a corresponding Monte Carlo error.

\subsection{Selecting summaries}
The choice of the summary statistic $S(\cdot)$ is paramount in any implementation of the ABC
methodology if one does not want to end up with simulations from the prior distribution resulting
from too large a tolerance! On the opposite, an efficient construction of $S(\cdots)$ may result in
a very efficient approximation for a given computational effort.

The literature on ABC abounds with more or less recommendable solutions to achieve a proper selection of the
summary statistic. Early studies were either experimental \citep{mckinley:cook:deardon:2009} or borrowing from
external perspectives. For instance, 
\citet{blum:francois:2010} argue in favour of using neural nets in their non-parametric modelling for
the very reason that neural nets eliminate irrelevant components of the summary statistic. However, the black
box features of neural nets also mean that the selection of the summary statistic is implicit. Another
illustration of the use of external assessments is the experiment ran by \citet{sedki:pudlo:2012} in mixing
local regression \citep{beaumont:zhang:balding:2002} local regression tools with the BIC criterion.

In my opinion, the most accomplished (if not ultimate) development in the ABC literature about the selection of the
summary statistic is currently found in \citet{fearnhead:prangle:2012}. Those authors study the use of a summary
statistic $S$ from a quasi-decision-theoretic perspective, evaluating the error by a quadratic loss
$$
L(\theta,d) = (\theta-d)^\text{T}A(\theta-d)\,,
$$
where $A$ is a positive symmetric matrix,
and obtaining in addition a determination of the optimal
bandwidth (or tolerance) $h$ from non-parametric evaluations of the error. In particular, the authors argue
that the optimal summary statistic is $\mathbb{E}[\theta|\bx^0]$ (when estimating the parameter of interest
$\theta$). For this, they notice that the errors resulting from an ABC modelling are of three types:
\begin{itemize}
\item[--] one due to the approximation of $\pi(\theta|\bx^0)$ by $\pi(\theta|\sobs)$,
\item[--] one due to the approximation of $\pi(\theta|\sobs)$ by
$$
\pi_\text{ABC}(\theta|\sobs) = \dfrac{
\int \pi(\bs) K[\{\bs-\sobs\}/h] \pi(\theta|\bs)\,\text{d}\bs
}{
\int \pi(\bs) K[\{\bs-\sobs\}/h]\,\text{d}\bs
}
$$
where $K(\cdot)$ is the kernel function used in the acceptance step---which is the indicator function
$\mathbb{I}_{(-1,1)}$ in the above algorithm since $\theta^{\star}$ is accepted with probability
$\mathbb{I}_{(-1,1)}(\rho(S(\mathbf{x}^0),S(\mathbf{x}^{*})/\epsilon)$ in this case---,
\item[--] one due to the approximation of
$\pi_\text{ABC}(\theta|\sobs)$ by importance Monte Carlo techniques based on $N$ simulations, which amounts to
$\text{var}(a(\theta)|\sobs)/N_\text{acc}$, if $N_\text{acc}$ is the expected number of acceptances.
\end{itemize}
For the specific case when $S(\bx)=\mathbb{E}[\theta|\bx]=\hat\theta$, the expected loss satisfies
$$
\mathbb{E}[L(\theta,\hat{\theta})|\bx^0]=\mbox{trace}(A\Upsigma)+h^2\int \bx^T A \bx K(\bx)\mbox{d}\bx+o(h^2)\,,
$$
where $\Upsigma=\mathrm{var}(\theta|\bx^0)$,
which means that the first type error vanishes with small $h$'s, given that it is equivalent to the Bayes risk
based on the whole data. From this decomposition of the risk, \citet{fearnhead:prangle:2012} derive
$$
h=O(N^{-1/(4+d)})
$$
as an optimal bandwidth for the standard ABC algorithm. From a practical perspective,
using the posterior expectation $\mathbb{E}[\theta|\bx^0]$ as a summary statistic is obviously impossible, if only
because even basic simulation from the posterior is impossible. \citet{fearnhead:prangle:2012} suggest using instead a
two-stage procedure:
\begin{enumerate}
\item Run a basic ABC algorithm to construct a non-parametric estimate of $\mathbb{E}[\theta|\bx^0]$ following
\citet{beaumont:zhang:balding:2002}; and
\item Use this non-parametric estimate as the summary statistic in a second ABC run.
\end{enumerate}
In cases when producing the reference sample is very costly, the same sample may be used in both runs,
even though this may induce biases that will simply add up to the many approximative steps inherent to
this procedure.

In conclusion, the literature on the topic has gathered several techniques
proposed for other methodologies. While this perspective manages to eliminate
the less relevant components of a pool of statistics, I feel the issue remains
quite open as to which statistic should be included at the start of an ABC algorithm. The
problems linked with the curse of dimensionality (``not too many"),
identifiability (``not too few"), and ultimately precision (``as many as components of $\theta$") of the
approximations are far from solved and I thus foresee further major developments to occur in the years to come.

\subsection{ABC model choice}\label{sec:modX}

As stressed already above, model choice occupies a special place in the
Bayesian paradigm and this for several reasons. First, the comparison of
several models compels the Bayesian modeller to construct a meta-model that
includes all these models under comparison as special cases. This encompassing
model thus has a complexity that is higher than the complexities of the models
under comparison. Second, while Bayesian inference on models is formally
straightforward, in that it computes the posterior probabilities of the models
under comparison---even though this raises misunderstanding and confusion in
the non-Bayesian applied communities, as illustrated by the series of
controversies raised by Templeton (\citeyear{templeton:2008,templeton:2010}---,
the computation of such objects often faces major computational challenges.

From an ABC perspective, the specificity of model selection holds as well. At
first sight, and in sort of predictable replication of the theoretical setting,
the formal simplicity of computing posterior probabilities can be mimicked by
an ABC-MC (for model choice) algorithm as the following one \citep{toni:stumpf:2010}:

\begin{algorithm}
\caption{ABC-MC}
\begin{algorithmic}
\FOR {$t=1$ to $T$}
\REPEAT
\STATE Generate  $m^{*}$ from the prior $\pi(\mathcal{M}=m)$.
\STATE Generate $\theta_{m^{*}}^{*}$ from the prior $\pi_{m^{*}}(\cdot)$.
\STATE Generate $x^{*}$ from the model $f_{m^*}(\cdot|\theta_{m^{*}}^{*})$.
\STATE Compute the distance $\rho(S(\mathbf{x}^0),S(\mathbf{x}^{*}))$.
\STATE Accept $(\theta_{m^{*}}^{*},m^{*})$ if $\rho(S(\mathbf{x}^0),S(\mathbf{x}^{*}))<\epsilon$.
\UNTIL acceptance
\ENDFOR
\end{algorithmic}
\end{algorithm}

where $\mathcal{M}$ denotes the unknown model index, $m$ being one of the possible values, 
with $\pi_m$ the corresponding prior on the parameter $\theta_m$.

\piccaption{\label{fig4}
Box-plots of the repartition of the ABC posterior probabilities that a normal {\em (Gauss)} and double-exponential
{\em (Laplace)} sample is from a normal (vs. double-exponential) distribution. based on 250 replications and
the median as summary statistic $S$ ({\em Source: \citealp{marin:pillai:robert:rousseau:2011})}.}
\piccaptioninside
\parpic[l]{\includegraphics[width=.6\textwidth]{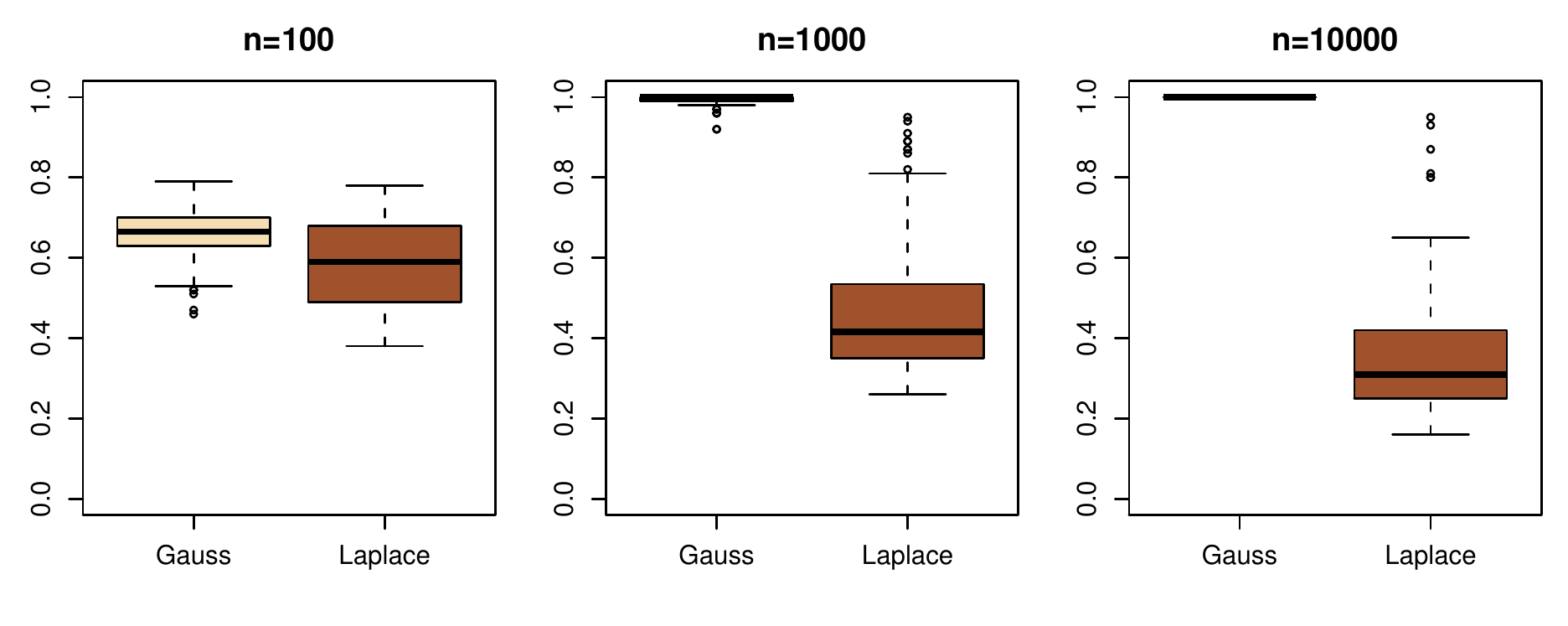}}

As a consequence, the above algorithm process the pair $(m,\allowbreak\theta_m)$ as a regular parameter, using the same tolerance
condition $\rho(S(\mathbf{x}^0),\allowbreak S(\mathbf{x}^{*}))\allowbreak <\epsilon$ 
as the initial ABC algorithm. From the output of
ABC-MC, the posterior probability $\pi(\mathcal{M}=m|\by)$ can then be approximated by the frequency of
acceptances of simulations from model $m$
$$ 
\hat\pi(\mathcal{M}=m|\by) = \dfrac{1}{T}\,\sum_{t=1}^T \mathbb{I}_{m^{(t)}=m}\,.  
$$
Improvements on this crude frequency estimate can be made using for instance a weighted polychotomous logistic
regression estimate of $\pi(\mathcal{M}=m|\by)$, with non-parametric kernel weights, as in
\citet{cornuet:santos:beaumont:etal:2008}.

\begin{example}[Example 1 (quinquies)]
If we resume our comparison of the normal and double-exponential models. Running ABC-MC in this case means
\begin{enumerate}
\item picking normal $m=1$ or double-exponential $m=2$ with probability $\nicefrac{1}{2}$;
\item simulating $\mu_m\sim\mathcal{N}(0,\sigma^2)$;
\item simulating a normal $\mathcal{N}(\mu_1,1)$ sample $\bx^*$ if $m=1$ and a double-exponential 
$\mathcal{L}(\mu_2,\nicefrac{1}{\sqrt{2}})$ sample $\bx^*$ if $m=2$;
\item compare $S(\bx^0)$ and $S(\bx^*)$
\end{enumerate}
While the choice of $S(\cdot)$ is unlimited, some choices are relevant and others are to be
avoided as discussed in \citet{robert:cornuet:marin:pillai:2011}. Figures \ref{fig4} and \ref{fig5}
show the difference in using for $S$ the median of the sample (Figure \ref{fig4}) and the median absolute
deviation (mad, defined as the median of the absolute values of the differences between the sample and its
median, $\text{med}(|x_i-\text{emd}(x_i)|)$) statistics (Figure \ref{fig5}). In the former case, double exponential
samples are not recognised as such and the posterior probabilities do not converge to zero. In the later case, they
do, which means the ABC Bayes factor is consistent in this setting.
\end{example}


\piccaption{\label{fig5}
Same legend as Fig.~\ref{fig4} when the summary statistic $S$ is the mad statistic
({\em Source: \citealp{marin:pillai:robert:rousseau:2011})}.}
\piccaptioninside
\parpic[r]{\includegraphics[width=.6\textwidth]{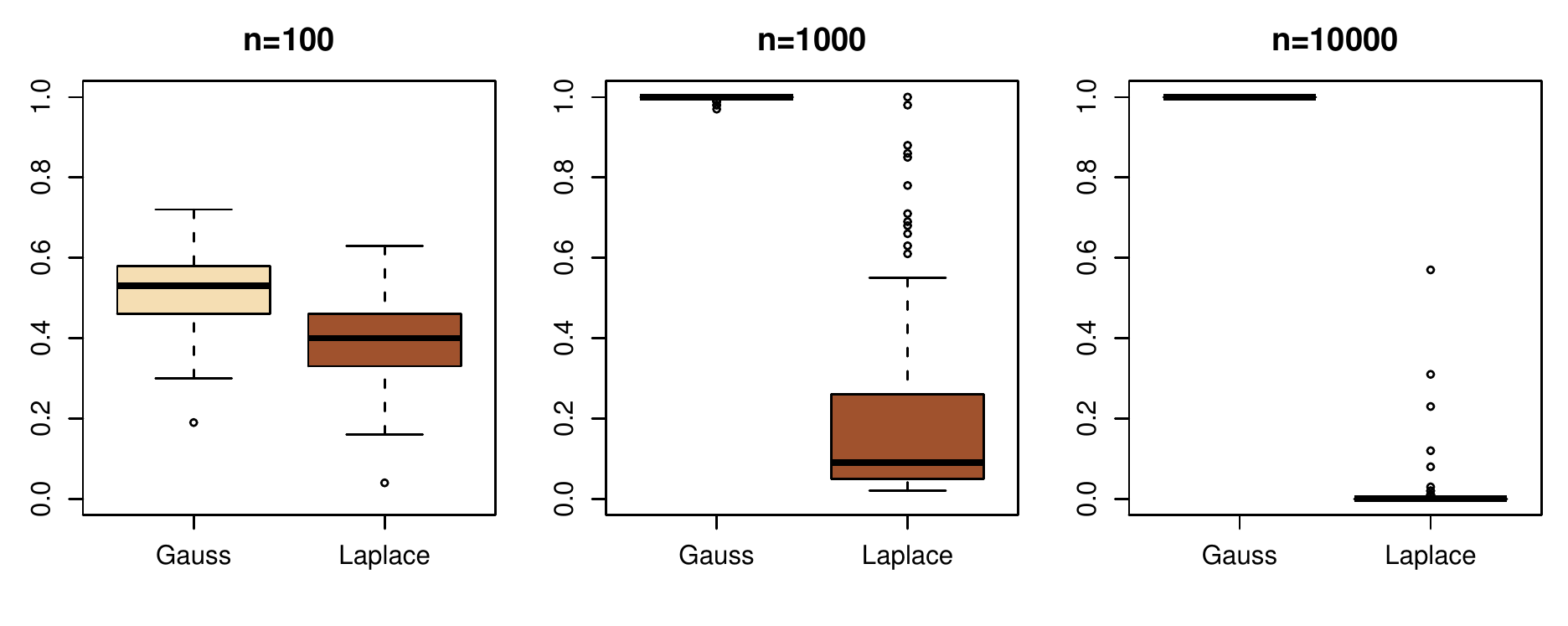}}

The conclusion of \citet{robert:cornuet:marin:pillai:2011} is that the outcome
of an ABC model choice based on a summary statistic that is insufficient may be
untrustworthy and need to be checked by additional Monte Carlo experiments as
those proposed in DIYABC \citep{cornuet:santos:beaumont:etal:2008}. More recently,
\citet{marin:pillai:robert:rousseau:2011} exhibited conditions on the summary statistic
for an ABC model choice approach to provide a consistent solution.

\section{Beyond}

This chapter provides a snapshot via a few illustrations of the diversity of
Bayesian computational techniques. It also misses important directions, like
the particle methods which are particularly suited for complex dynamical models
\citep{delmoral:doucet:jasra:2006,andrieu:doucet:holenstein:2010}. Or
variational Bayes techniques which rely on optimised approximations to a
complex target distribution \citep{jaakkola:jordan:2000}. Or partly analytical
integration taking advantage of Gaussian structures, as for the quickly expanding
INLA technology \citep{rue:martino:chopin:2009}, which recent advances are covered
by \citet{martins:simpson:lindgren:rue:2013}. Or yet more remote
approximations to the likelihood function based on higher order asymptotics
\citep{ventura:cabras:racugno:2009}. Similarly, I did not mention recent
simulations methodologies that coped with non-parametric Bayesian problems
\citep{hjort:holmes:mueller:walker:2010} and with stochastic processes
\citep{beskos:papaspiliopoulos:roberts:fearnhead:2006}. The field is expanding
and the demands made by the ``Big Data" crisis are simultaneously threatening
the fundamentals of the Bayesian approach by calling for quick-and-dirty
solutions and bringing new materials, by exhibiting a crucial need for
hierarchical Bayes modelling. Thus, to conclude with Dickens'
(\citeyear{dickens:1859}) opening words, we may later consider that ``it was
the best of times, it was the worst of times, it was the age of wisdom, it was
the age of foolishness".

\section*{Acknowledgements}

I am quite grateful to Jean-Michel Marin for providing some of the material
included in this chapter, around Example 4 and the associated figures. It
should have been part of the chapter on hierarchical models in our new book
{\em Bayesian essentials with R}, chapter that we eventually had to abandon to
its semi-baked status. The section on ABC was also salvaged from another
attempt at a joint survey for a Statistics and Biology handbook, survey that
did not evolve much further than my initial notes and obviously did not meet
the deadline.  Therefore, Jean-Michel should have been a co-author of this
chapter but he repeatedly declined my requests to join. He is thus named
co-author {\em in absentia}. Thanks to Jean-Louis Foulley, as well, who
suggested using the Pothoff and Roy (1964) dataset in his ENSAI lecture notes.


\end{document}